# Nonreciprocal quantum photon-pair source with chiral ferroelectric nematics

Jin-Tao Pan[1,6], Yun-Kun Wu[2,3,4,6], Ling-Ling Ma[1,6,*], Ning Wang[1], Xin-Yu Tao[1], Bo-Han Zhu[1], Shu Wang[2,3,4], Fang-Wen Sun[2,3,4], Guang-Can Guo[2], Hui Jing[5,*], Xi-Feng Ren[2,3,4,*] & Yan-Qing Lu[1,*]

[1]National Laboratory of Solid State Microstructures, Key Laboratory of Intelligent Optical Sensing and Manipulation, Collaborative Innovation Center of Advanced Microstructures, College of Engineering and Applied Sciences, Nanjing University, Nanjing, China.

[2]CAS Key Laboratory of Quantum Information, University of Science and Technology of China, Hefei, China.

[3]CAS Synergetic Innovation Center of Quantum Information and Quantum Physics, University of Science and Technology of China, Hefei, China.

[4]Hefei National Laboratory, University of Science and Technology of China, Hefei, China.

[5]Key Laboratory of Low-Dimensional Quantum Structures and Quantum Control of Ministry of Education, Department of Physics and Synergetic Innovation Center for Quantum Effects and Applications, Hunan Normal University, Changsha, China.

[6]These authors contributed equally: Jin-Tao Pan, Yun-Kun Wu, Ling-Ling Ma.

[*]e-mail: malingling@nju.edu.cn; jinghui@hunnu.edu.cn; renxf@ustc.edu.cn; yqlu@nju.edu.cn.

## Abstract

Quantum nonreciprocity—a fundamental phenomenon enabling directional control of quantum states and photon correlations—has long been recognized as pivotal for quantum technologies. However, the experimental realization of nonreciprocal quantum photon-pair generation, as a critical prerequisite for advancing quantum systems, continues to be an outstanding challenge that remains unaddressed in practice. Here, we experimentally implement a highly-efficient nonreciprocal quantum photon source in a micro/nano-scale helical structured nonlinear optical fluid. Intriguing helical quasi-phase matching is achieved by deliberately engineering the pitch of the chiral ferroelectric structure, thus enabling spontaneous parametric down-conversion with record-high brightness (5,801.6 Hz·mW$^{-1}$, 10,071% enhancement over phase-mismatched systems) and high coincidence-to-accidental ratio, rivaling state-of-the-art centimeter-scale nonlinear crystals. In particular, by tailoring the ferroelectric helix structure with orthogonally aligned head and tail polarization vectors, we demonstrate up to 22.6 dB isolation in biphoton generation coupled with **nonreciprocal** quantum polarization states, while maintaining classical optical **reciprocity**. This quantum liquid-crystal-based platform, combining flexible tunability and superior performance of purely quantum nonreciprocity at micro/nano scales, builds a bridge between a wide range of soft-matter systems, nonreciprocal physics, and emerging quantum photonic technologies.

# Main

Nonreciprocal devices, featuring distinct responses when the source and receiver are exchanged, as already fabricated in diverse fields ranging from electronics or photonics[1] to acoustic or thermal materials[2], have played indispensable roles in many important applications, such as isolators or circulators widely used in modern optical communications and emerging quantum techniques[3]. Very recently, rapid advances have been witnessed in achieving a one-way flow of single photons in purely quantum systems[4-21], which are crucial for making backaction-immune quantum chips or building directional quantum networks. However, previous works have mainly focused on nonreciprocal control of classical mean transmission rates instead of any quantum fluctuations or high-order correlations of photons. The possibilities of achieving nonreciprocal quantum effects, such as one-way single-photon blockade[22-27] and directional quantum entanglement[28-31], have been predicted, which have no classical counterparts at all and can appear even when the systems are reciprocal at the classical level, i.e., the transmission rates are the same while the quantum correlations are qualitatively different at different output ports. Some unique effects of quantum nonreciprocity have been verified in experiments using a purely optical cavity[32] or a cavity-QED system[33]. Despite these pioneering advances, nonreciprocal quantum entangled photon pairs, as essential resources of quantum techniques, have not been realized experimentally, hindering their applications, e.g., nonreciprocal quantum metrology[34].

Here, we propose and demonstrate how to achieve this goal by using liquid crystals (LCs). LCs featuring unique self-assembly properties, flexible electric tunability, and high sensitivity to external signals have emerged as a promising platform for photoelectric applications[35-37]. The recent discovery of ferroelectric nematic LCs (FNLCs) has further expanded their potential due to their inherent second-order optical nonlinearity[38-43]. Very recently, in LC systems, the implementation of spontaneous parametric down-conversion (SPDC), a fundamental process for generating entangled photons, has been reported[44]. In this work, we show that by combining the nonlinear SPDC and the controllable breaking of space-reversal symmetry in self-assembled chiral helical ferroelectric architectures, nonreciprocal quantum entangled photon sources can be realized with LC devices. The nonreciprocal generation of bright, broadband entangled photon pairs, as first demonstrated here, can serve as a promising soft-matter platform for exploring nonreciprocal quantum physics and applications in quantum information techniques.

Specifically, in order to break reciprocity in LCs, we incorporate chirality into the self-assembled FNLCs to induce a rationally designed chiral ferroelectric structure. For a 50 μm-thick LC device, the measured brightness reaches 5801.6 Hz·mW$^{−1}$, surpassing traditional non-phase-matched configurations by an impressive factor of 10071%. This brightness and the coincidence-to-accidental ratio are comparable to state-of-the-art centimeter-scale beta barium borate (BBO) crystals. By tuning the ferroelectric LC helixes with orthogonal orientations of head and tail LC directors, our platform further demonstrates marked quantum nonreciprocity in the SPDC process, while the linear optical processes retain reciprocity. We achieve an isolation ratio of up to 22.6 dB, representing a significant advancement in realizing high-fidelity nonreciprocal quantum devices. This work not only offers a compact and scalable solution for exploring nonreciprocal quantum photonics but also opens new avenues for realizing next-generation noise-tolerant quantum communication, backaction-immune quantum networks, and one-way quantum sensing.

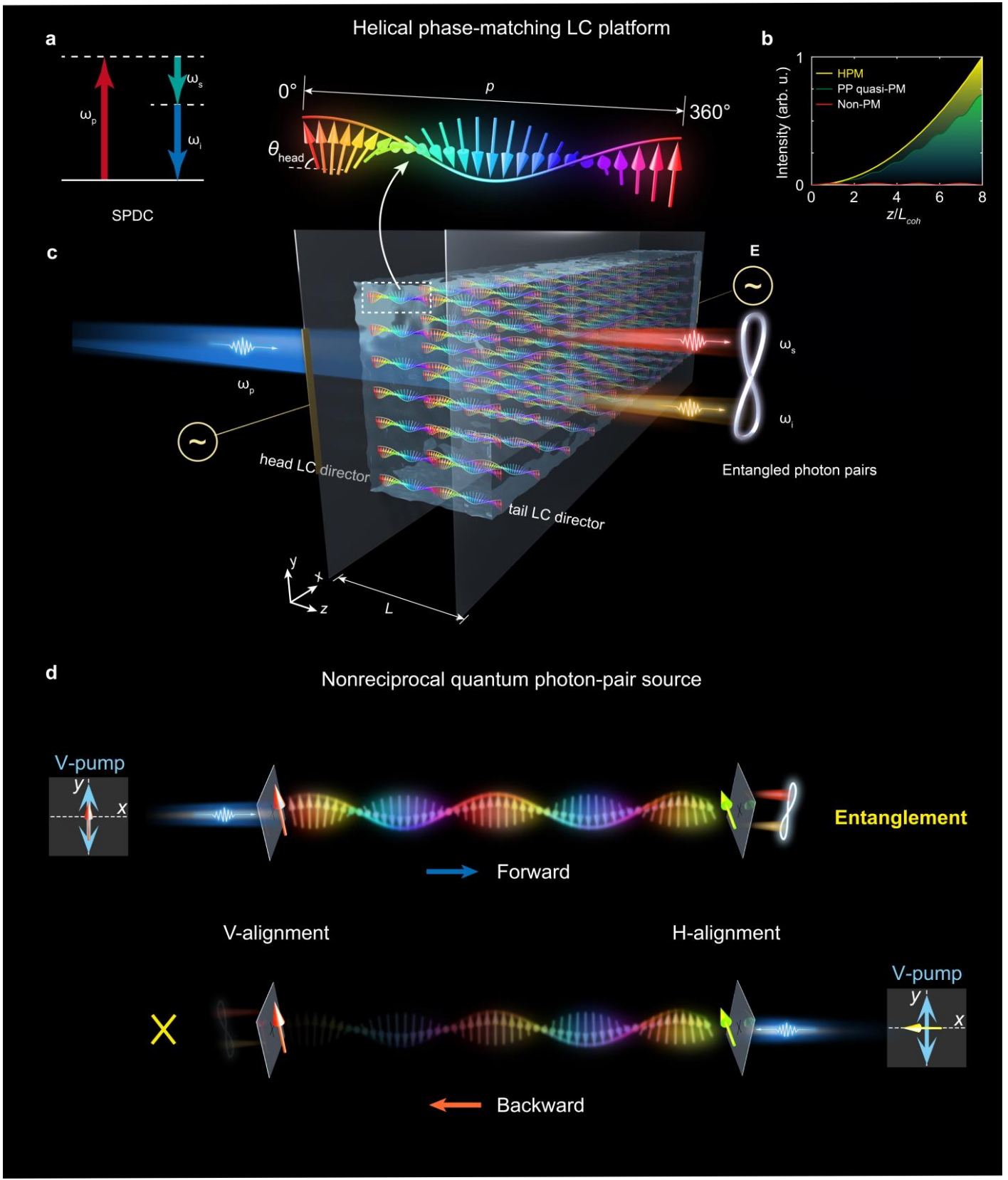

**Figure 1 | Schematic of the SPDC process in helical ferroelectric fluids. a**, Illustration of the energy conservation condition in the SPDC process. **b**, Variation of intensity along the nonlinear medium for helical quasi-phase matching, periodically poled quasi-phase matching, and non-phase matching cases. HPM: helical quasi-phase matching; PP quasi-PM: periodically poled quasi-phase matching; Non-PM: non-phase matching. **c**, Generation of entangled photon pairs with a helical FNLC. For the FNLC doped with chiral molecules R811, the self-assembly periodic spontaneous polarization acts as the nonlinear phase compensation structure. **d**, Schematic diagram of nonreciprocal SPDC in orthogonally aligned chiral phase-matched FNLC.

## Principle

SPDC, a cornerstone nonlinear optical process in which a high-energy photon splits into two lower-energy photons naturally entangled in the temporal and frequency domains **(Figs. 1a-c)**, serves as the primary mechanism for generating entangled photon pairs.

Here, we engineer a rationally designed chiral ferroelectric nematic structure to efficiently modulate the SPDC quantum nonlinear process. Within this architecture, the polar LC director field $\vec{n}(z)$ exhibits a continuous sinusoidal modulation along the propagation direction (*z*):

$$\vec{n}(z) = \begin{pmatrix} \cos(Gz + \theta_{\text{head}}) \\ \sin(Gz + \theta_{\text{head}}) \end{pmatrix}, \quad (1)$$

where $\theta_{\text{head}}$ is the orientation angle of the head LC director of the helixes at the incident surface, $G = 2\pi/p$ is the reciprocal vector determined by the helical pitch (*p*).

For linearly polarized pump light, we simulate the polarization evolution during its propagation through the LC medium and find that the inherent optical rotatory effect of such a structure induces an adiabatic polarization rotation that aligns with the helical director orientations[45]. Although, in our practical case, the linear polarization slightly converts to elliptical polarization (**Fig. S1**), it predominantly retains the rotation rules mentioned above (**SI Section 1**). Based on this observation, we approximate the LC system under the Mauguin condition ($\Delta n \cdot p \gg \lambda$, where $\lambda$ is the wavelength of the pumping light, $\Delta n$ is the birefringence). Then, the entangled photon-pair generation rate via SPDC is simplified as follows (**SI Section 2**):

$$R_{\text{SPDC}} \propto P_{\text{pump}} L^2 \text{sinc}^2\left(\frac{(\Delta k - G)L}{2}\right) \cdot d_{\text{eff}}^2 = P_{\text{pump}} L^2 \text{sinc}^2\left(\frac{\Delta k L}{2} - \pi\frac{L}{p}\right) \cdot d_{33}^2 \cdot \cos^2(\theta_{\text{head}} - \theta_{\text{pol}}), \quad (2)$$

where $d_{33}$ is the dominant second-order nonlinear susceptibility of LC[46], $P_{\text{pump}}$ is the pump power, $\Delta k$ ($\Delta k = k_p - k_s - k_i \neq 0$, where $k_p$, $k_s$, and $k_i$ denote the wavevectors of the pump, signal, and idler fields, respectively) represents the wavevector mismatch, and *L* is the LC film thickness. This expression incorporates two modulation terms: (i) $\text{sinc}^2$ and (ii) $\cos^2$. Obviously, the first term is controlled via *p* and can be optimized by $\frac{\Delta kL}{2} - \pi\frac{L}{p} = 0$, thus enabling a helical quasi-phase matching (HPM) strategy to enhance the efficiency of the entangled photon-pair generation essentially. Once the HPM is satisfied, the entangled biphoton coincidence rate will scale quadratically with *L* (**Fig. 1b**), demonstrating thickness-dependent efficiency scaling for specific linearly polarized pumping light.

Furthermore, we can see from the second term that the SPDC emission efficiency is also governed by the relationship between $\theta_{\text{pol}}$ and $\theta_{\text{head}}$. Specifically, the maximum generation rates ($R_{\text{max}}$) occur when the incident linear polarization aligns parallel to the polar LC director at the light entry interface, whereas an orthogonal director orientation yields minimal generation ($R_{\text{min}}$). For a specific linear polarization incidence, we can tailor $\theta_{\text{head}}$ to tune the generation rate of entangled photon pairs. Consider the scenario where we simultaneously engineer both the head and tail LC director orientations ($\theta_{\text{head}}$ and $\theta_{\text{tail}}$) such that they are mutually perpendicular. This configuration, for the incident linear polarization parallel to the head/tail LC director, would elicit distinct quantum responses in the SPDC process for forward and backward propagation (**Fig. 1d**). This phenomenon could potentially enable the realization of nonreciprocal entangled photon-pair generation, thereby unlocking new avenues in quantum optics and photonics.

## Phase-matching SPDC source of entangled photon pairs

To validate the HPM scheme, we engineered a chiral FNLC system, where the chiral molecular dopants enable precise control over the helical architecture. A unidirectionally photoaligned LC sample with optimized helical pitch was fabricated, featuring a polar LC helix structure with both the head and tail director orientations along the *y*-axis. For continuous-wave pumping light at 650 nm with *y*-axis linear polarization, the HPM LC sample demonstrates a record-high SPDC brightness of $3.3 \times 10^4$ pairs/s—a 10,071% enhancement over non-phase-matched counterparts of identical thickness (**Fig. 2a**). This dramatic improvement aligns with our theoretical calculation (**SI Section 2**).

The second-order correlation function $g^{(2)}$ is characterized, revealing an ultrahigh and unambiguous coincidence peak at zero delay time **(Fig. 2b)**. This observation confirms the presence of pronounced temporal quantum correlations between the generated photon pairs, with minimal noise contributions from classical optical processes. The spectral property of the entangled photon pairs, as shown in **Fig. 2c**, is measured using two-photon temporally correlated fiber spectroscopy[47] and exhibits a remarkable broadband profile spanning ~150 nm. To some extent, this spectral bandwidth remains constrained by the operational bandwidth of our detector, which is anticipated to be broader through optimized detector selection. This broadband nature is primarily governed by the interaction length, material dispersion characteristics, and HPM configuration. These results demonstrate an optimal balance between efficient entangled photon-pair generation and robust phase-matching conditions, ensuring both high yield and stability in the SPDC process.

Then, we systematically investigate the polarization-dependent coincidence rate of generated photon pairs, **Fig. 2d**. A polarization control system comprising a linear polarizer (LP) and a quarter-wave plate (QWP) enables continuous tuning of the incident polarization state, spanning linear, circular, orthogonal linear, and opposite circular polarizations. Initial polarization states are configured as vertical (V, red trace) and horizontal (H, blue trace) linear polarizations. Spatial analysis of coincidence rates across four cardinal orientations revealed that V-polarized pumping light consistently generates higher photon-pair brightness, of which the polarization is parallel to the head LC director. Under circularly polarized excitation, the intrinsic chirality of the ferroelectric fluidic structure introduces a distinct circular dichroism (CD) effect into the SPDC process. Specifically, at azimuthal angles of 45°, 135°, 225°, and 315° in this figure, left-handed circularly polarized pumping, i.e., chirality-opposed to the structure helicity, demonstrates enhanced coincidence rates ($R_{\text{LCP}}$), while the right-handed, syn-chirality circular polarization induces low coincidence rates ($R_{\text{RCP}}$). By defining the parameter of $\text{CD}_{\text{SPDC}} = (R_{\text{LCP}} - R_{\text{RCP}})/(R_{\text{LCP}} + R_{\text{RCP}})$, we obtain $\text{CD}_{\text{SPDC}} = 0.54$. Notably, the observed CD phenomenon is rarely manifested in conventional nonlinear quantum systems. This phenomenon provides compelling evidence for the critical role of mirror symmetry breaking arising from the intrinsic chiral architecture of LCs, thereby enabling polarization-handedness-dependent quantum state engineering in entangled photon-pair generation. Theoretical simulations, represented by dotted traces, demonstrated an excellent agreement with experimental observations, validating the proposed SPDC source mechanism. More detailed polarization dependences of the coincidence rate of generated photon pairs are provided in **Fig. S2**.

To further evaluate the performance of the SPDC process, we conduct measurements of the coincidence rate and the coincidence-to-accidental ratio (CAR) depending on pumping power and LC film thickness. **Fig. 2e** showcases a characteristic linear increase in coincidence rate with pumping power and an inverse proportionality trend of CAR, both consistent with SPDC theory. The coincidence rate reaching 5,801.6 Hz·mW$^{-1}$ alongside a high CAR signifies a micro-scale SPDC source whose performances rival that of commercial centimeter-scale bulk BBO crystals. Such high CAR values particularly highlight the efficient entangled photon-pair generation capability with very low noise, underscoring the advantages of our chiral ferroelectric nematic platform for miniaturized quantum light sources. The thickness-dependent biphoton coincidence rate is shown in **Fig. 2f**. By incrementally increasing the LC film thickness from 8 μm to 50 μm—far exceeding the calculated coherent length ($L_c = \pi/\Delta k = 15$ μm)—while maintaining a fixed helical pitch, we observe that the coincidence rate follows a distinct quadratic dependence on the LC thickness ($R \propto L^2$), aligning precisely with our theoretical prediction for the HPM SPDC source. This scaling behavior fundamentally differs from two established cases. One is the non-phase-matching system exhibiting oscillatory intensity profiles due to periodic destructive interference. The other is the conventional quasi-phase matching (QPM) configuration (e.g., periodically poled nonlinear crystals[48]), which achieves discrete phase-matching through periodic polar domain inversions, thus manifesting step-like intensity enhancements at specific poling periods (**Fig. 2f**). The continuous LC director rotation in our HPM helical ferroelectric LC architecture directly demonstrates its superior capability for achieving higher photon-pair brightness over extended interaction lengths due to better Fourier purity and lower loss (**SI Section 3**). Importantly, these performance improvements are achieved without the stringent fabrication tolerances or elevated costs of conventional QPM techniques. Instead, the chiral ferroelectric nematic architecture is realized through self-assembly processes, which offer distinct advantages, including simplified fabrications, dynamic reconfigurability, cost-effectiveness, and energy-efficient processing.

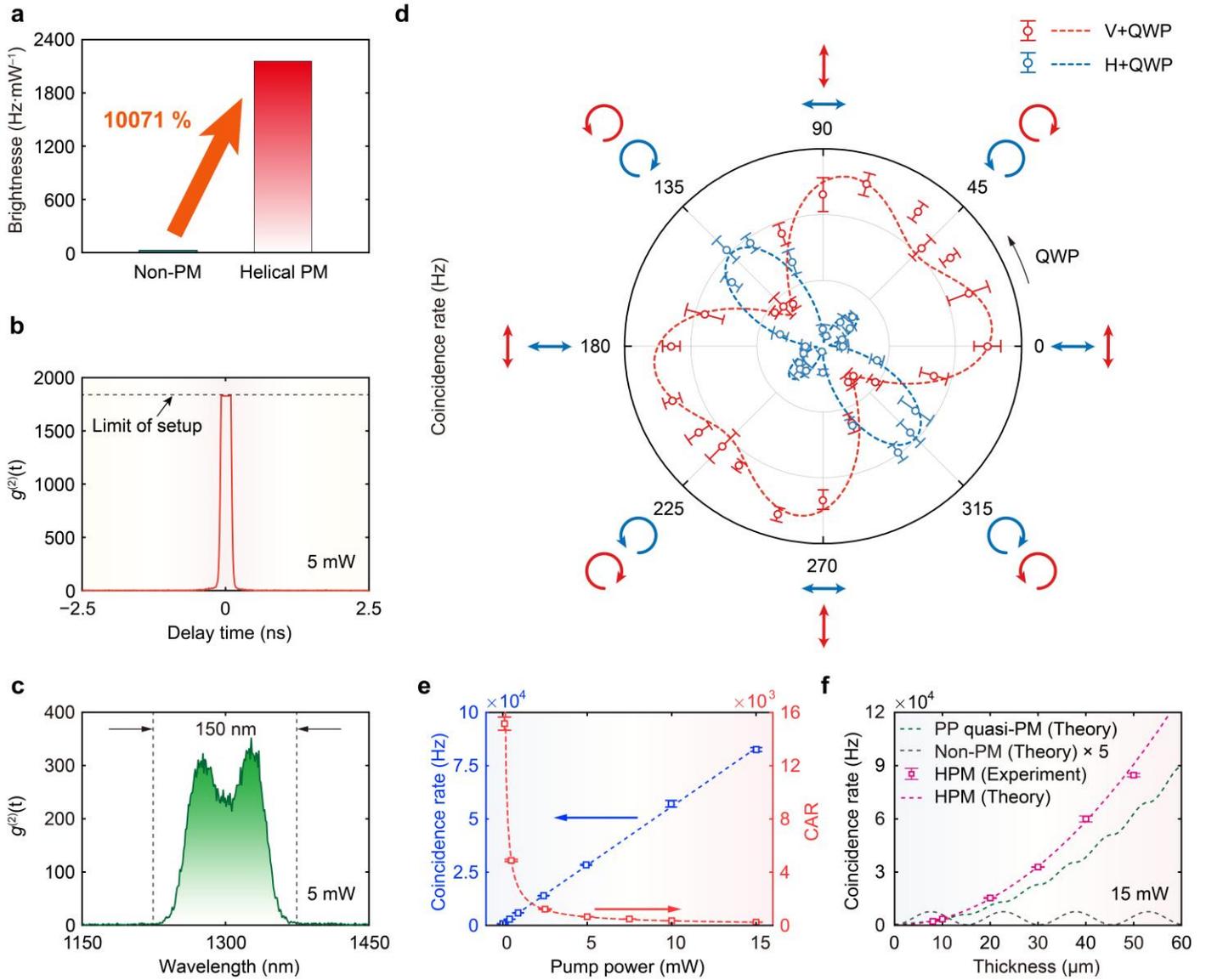

**Figure 2. Phase-matching enhanced spontaneous parametric down-conversion in chiral LC**. **a**, Brightness comparison between non-phase-matched and helical phase-matched SPDC. **b**, Two-photon temporal correlation function in a helical phase-matched LC under ∼5 mW pump power from a 650 nm continuous-wave laser. **c**, Broadband spectrum of generated photon pairs (limited by the cutoff wavelength of filters and detection bandwidth, ∼150 nm). **d**, Coincidence rates versus pump polarization. **e**, Linear increase of biphoton coincidence rate with pump power. The CAR exhibits an inverse proportionality to pump power, confirming biphoton emission. **f**, Thickness-dependent coincidence rate at 15 mW pump power. Error bars represent standard deviations from multiple measurements.

## Nonreciprocal quantum photon-pair generation

We report the pronounced quantum nonreciprocity in SPDC processes while preserving classical optical reciprocity, as illustrated in **Figs. 3a and 3b**, by engineering chiral ferroelectric helix structures with orthogonally aligned head and tail LC director orientations ($\theta_{\text{tail}} - \theta_{\text{head}} = 2\pi L/p$ equal to an odd multiple of $\pi/2$, **Fig. 1d**).

First, the reciprocity of classical light transmission is examined in such an asymmetric chiral ferroelectric architecture. As shown in **Figs. 3a and 3c**, the birefringence-driven optical rotatory effect inherent to this chiral architecture preserves time-reversal symmetry – despite the orthogonal LC director orientations at the structure's head and tail – as confirmed by the near-identical transmission efficiencies (**Fig. 3c**) and symmetric polarization state transformations (**Fig. S3**) for forward and backward propagations. This symmetry persistence aligns with the established principle of linear optics in chiral LC media[45], fundamentally rooted in the time-symmetric nature of Maxwell's equations.

In contrast, the chiral ferroelectric nematic configuration introduces a directional asymmetry in quantum state generation through controlled spatial modulation of the effective second-order nonlinear susceptibility ($\chi^{(2)}$). As demonstrated in **Figs. 3b and 3d**, the V-polarized pumping achieves a much higher coincident rate during the forward propagation compared to backward propagation. This is because, for the forward direction, the head LC director orientation is parallel to the incident polarization, thus enabling sustained constructive interference across the interaction length, according to Eq. (2). Conversely, under backward propagation with identical V-polarized incidence, the tail LC director orientation – now orthogonal to the incident polarization at the entry interface – creates a persistent geometric mismatch – resulting in significantly suppressed coincident rate. Consequently, a unidirectional entangled photon-pair emission with markedly different SPDC efficiencies is observed as the pumping power increases (**Fig. 3d**). Importantly, while the optical rotatory effect persists bidirectionally (the same as in classical regiem), the asymmetric helix-locked correlation between light polarization and material polarization imposes direction-dependent nonlinear coupling. This highlights the system's unique ability to break Lorentz reciprocity exclusively in the quantum regime.

To quantitative characterize the nonreciprocal quantum photon-pair generation, we define the nonreciprocal isolation ratio as

$$\text{Isotation (dB)} = 10\log_{10}(R_{\text{forward}}/R_{\text{backward}}), \tag{3}$$

aligning with standardized methodologies (**SI Section 4**) for quantifying the nonreciprocity, where $R$ denotes either the SPDC coincidence rate or the classical light transmission. The measured power-dependent isolations in both the classical linear realm and the quantum nonlinear regime are shown in **Fig. S4**, further demonstrating the pronounced quantum nonreciprocity while the linear optical process remains reciprocal. Interestingly and importantly, the SPDC source of entangled photon pairs based on our HPM chiral ferroelectric LCs exhibits power-independent nonreciprocal isolation across input powers (0.1~15 mW), which is different from many conventional nonreciprocal optical systems that exhibit power-dependent behavior[49,50]. This result promises robust, nonreciprocal quantum sources with very low powers (100 μW).

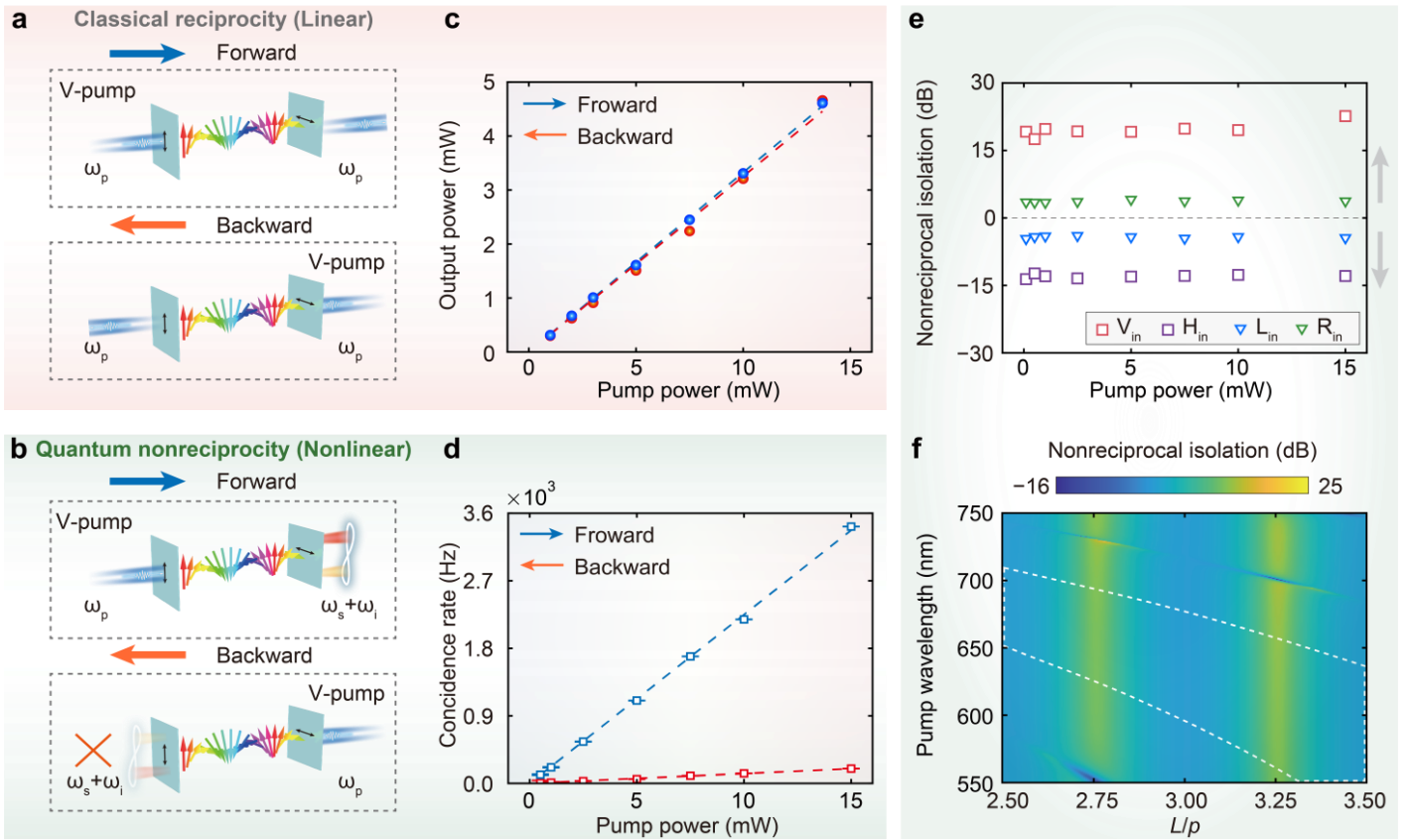

**Figure 3 | Reciprocal classical optics and nonreciprocal quantum SPDC in orthogonally aligned LC. a,b**, Schematic diagrams of (**a**) classical reciprocal light propagation and (**b**) quantum nonreciprocal SPDC processes in orthogonally aligned LC samples. **c**, Forward and backward transmitted power at 650 nm as a function of input power, demonstrating the linear reciprocal behavior. **d**, Forward and backward SPDC coincidence rates versus pump power, showing nonreciprocal characteristics. **e**, Experimental isolation contrast for nonreciprocal SPDC under vertically (V), horizontally (H), left-handed circularly (L), and right-handed circularly (R) polarized pumping. **f**, Calculated isolation contrast as a function of the thickness-to-pitch ratio ($L/p$) and pump wavelength. Notably, discrete singularity lines emerge in isolation ratios at complete phase mismatch conditions. These singularity isolation ratios are irrelevant since coincidence rates in both directions are negligible at those points.

In addition to the linear polarization excitation, a nonreciprocal SPDC response is also observed for circular polarization pumping due to the intrinsic circular dichroism of the HPM chiral LC structure. **Fig. 3e** presents the nonreciprocal isolation of entangled photon-pair generation under various pumping polarization configurations for a left-handed chiral ferroelectric device. We can see that the isolations are about $\pm 4$ dB for circularly polarized pumping, which is much lower than the values for V- and H-polarized pumping (highest to ~22.6 dB—the highest isolation ratio for quantum nonreciprocity to the best of our knowledge). This pronounced polarization-dependent nonreciprocity highlights the critical roles of both the head-tail alignment symmetry and the structural chirality in the ferroelectric LC architecture. The diminished isolation under circular polarization arises from the dynamic polarization evolution during propagation, specifically from circular, elliptical polarization to linear, reverse-handed elliptical polarization, and so on. This time-varying polarization profile couples weakly to local polar LC directors within the helical LC framework, thereby hindering the directional asymmetry. Conversely, V/H-polarized light aligns with either head or tail LC director orientations. This alignment facilitates strong coupling with the polar LC directors at every location through optical rotatory effects, thus effectively maximizing forward/backward asymmetry. More details can be found in **Figs. S5 and S6**.

To systematically engineer the nonreciprocal SPDC emission of our system, we numerically investigate the isolation ratio in a chiral ferroelectric nematic device with adjustable structural parameters. In this model, the pumping polarization and the head LC director orientation are both on the *y*-axis while maintaining a constant sample thickness, with the helical pitch (*p*) and pumping wavelength (*λ*) serving as independent variables (**Fig. 3f**). It is worth noting that the pitch variation will induce a concurrent modification in both momentum compensation ($\Delta k$) and the tail LC director orientation (with fixed head LC director orientation). The maximum isolation occurs at the thickness-to-pitch ratios (*L*/*p*) corresponding to odd-integer multiples of 0.25.

**Figs. 4a and 4b** further incorporate the HPM condition into the nonreciprocal SPDC emission under forward and backward pumping configurations. The LC helix structure exhibits orthogonally aligned head and tail LC director orientations, the same as that illustrated in **Fig. 3b**. Due to the parallel material polarization and incident light polarization, the coincidence rate in forward propagation (**Fig. 4a**) is governed solely by the HPM condition, whereas backward propagation (**Fig. 4b**) exhibits a dual dependence on both the HPM and the relationship between incident polarization and tail LC director. This mechanistic dichotomy suggests that the nonreciprocal isolation ratio basically relies on the orthogonal head-tail LC director configuration and is weakly affected by the phase-matching condition. At some special regions where the phase is completely mismatched, singularity lines of the nonreciprocal isolation ratio occur. But the isolation ratios in these regions are meaningless since there are almost no coincidence rates in both propagation directions. The white dashed contours in **Fig. 3f** indicate phase-matching regions (**Fig. 4a**) that exhibit a considerable overlap with the regions of high isolation ratio. Therefore, the high entangled photon-pair brightness and exceptional quantum nonreciprocity can be easily achieved simultaneously—a critical advancement experimentally and theoretically validated in our chiral ferroelectric LC platform.

Moreover, we determine the quantum polarization states of the generated entangled photon pairs, another significant physical quantity during the SPDC process. The experimental and corresponding theoretical polarization density matrices (*ρ*), which contain the complete information about the two-photon polarization states, are shown in **Figs. 4c and 4d**, for forward and backward propagations, respectively. The experimental density matrices are measured and reconstructed by the standard quantum tomography methods[51], as follows:

$$\begin{aligned}\psi_{\text{Forward}} &= 0.79e^{-2.69i}|\text{HH}\rangle + 0.39e^{1.88i}|\text{HV}\rangle + 0.42e^{1.72i}|\text{VH}\rangle + 0.23|\text{VV}\rangle \\ \psi_{\text{Backward}} &= 0.24e^{0.13i}|\text{HH}\rangle + 0.45e^{1.65i}|\text{HV}\rangle + 0.43e^{1.63i}|\text{VH}\rangle + 0.75|\text{VV}\rangle\end{aligned} \quad (4)$$

The corresponding theoretical density matrices are calculated as detail described in the **SI Section 5**. The forward and backward quantum polarization states in the experiment exhibit fidelities of $89 \pm 2\%$ and $93 \pm 1\%$ to the theoretical ones, respectively. Comparing these states, we can find an exchange of the maximum polarization components between $|\text{HH}\rangle$ and $|\text{VV}\rangle$. These results indicate that the quantum polarization states of the generated photon pairs in our proposal HPM LC system are also nonreciprocal, which contrasts with the behavior observed in the classical linear optical process. Interestingly, the quantum polarization states of the generated entangled photon pairs primarily depend on the tail LC director orientation at the exit surface, regardless of the head director orientation at the incident surface and the pumping polarization. This is further confirmed by the measurements in **Fig. S7**, which provide the quantum polarization states under different pump light polarization (H-polarization and diagonal polarization) for the same HPM LC device. By modulating the alignment direction or using electric tuning at the exit surface, the dominant component of the quantum polarization state, superposed by slight cross-polarization components of $|\text{HV}\rangle$ and $|\text{VH}\rangle$ states, is always parallel to the tail LC director orientation of the helix (see **Fig. S8**).

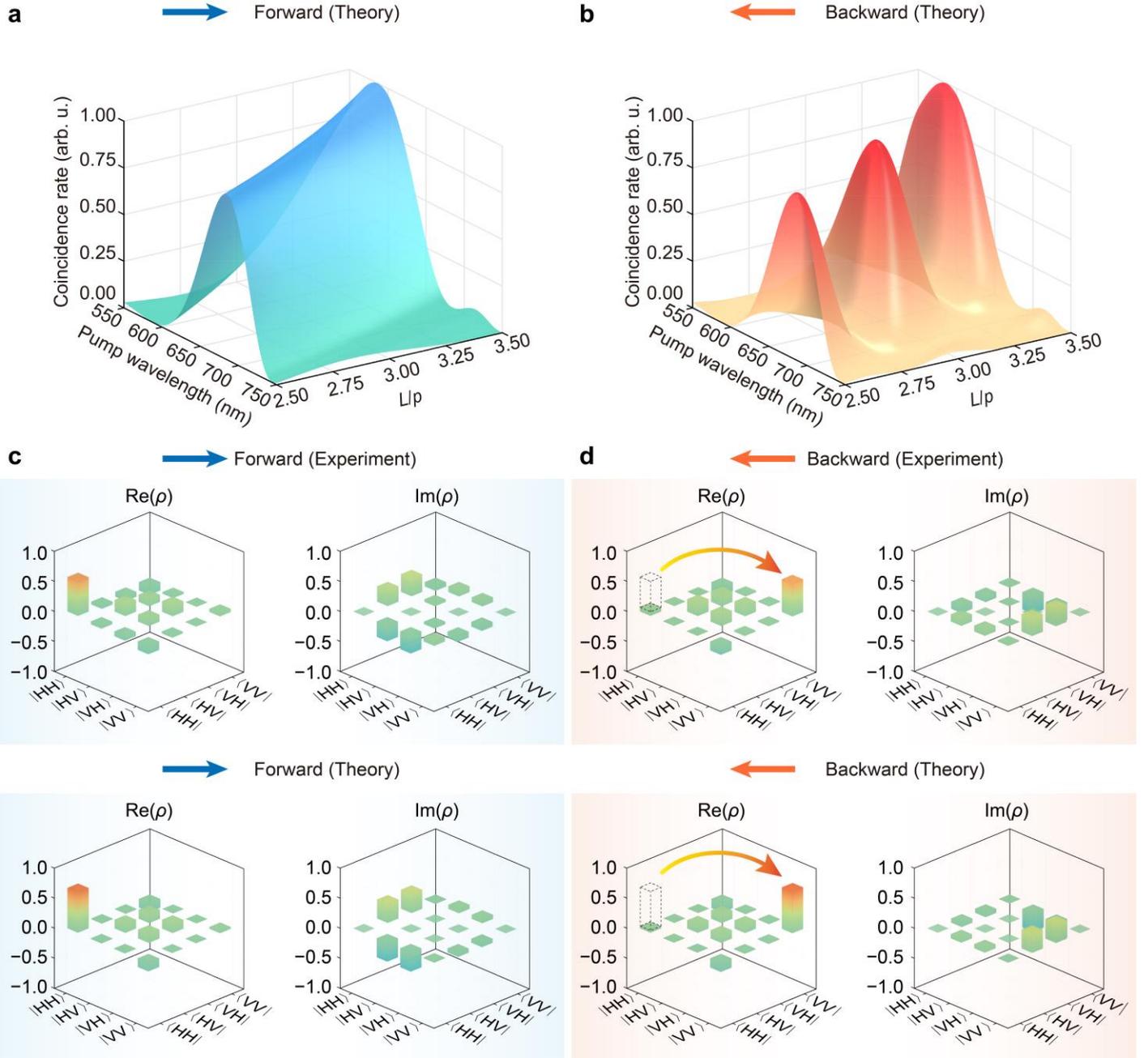

**Figure 4 | Nonreciprocal quantum polarization states in orthogonally aligned LCs. a,b**, Calculated SPDC coincidence rate map as a function of $L/p$ and pump wavelength for (**a**) forward and (**b**) backward pump, highlighting regions of enhanced phase-matching efficiency. **c,d**, Experimental (top) and theoretical (bottom) polarization density matrices ($\rho$) for forward and backward entangled photon pairs under V-polarized pumping.

## Discussion

In this study, we have successfully demonstrated a novel experimental realization of a nonreciprocal quantum photon-pair source based on a chiral soft matter platform. This platform leverages the unique properties of chiral ferroelectric helix structures, offering a rich tunable parameter space that includes the head and tail LC director orientations as well as the helical pitch. By strategically adjusting these parameters, we achieve significant advancements in both the nonreciprocity and efficiency of quantum photon-pair generation. An unprecedented isolation ratio of 22.6 dB is demonstrated in entangled biphoton generation, coupled with nonreciprocal quantum polarization states, even when the systems are reciprocal at the classical level. This study of nonreciprocal quantum light sources is crucial for building directional quantum networks or making backaction-immune quantum chips, and constitutes a pivotal frontier in contemporary physics, which not only addresses pressing challenges in quantum engineering but also deepens our understanding of symmetry breaking, topology, and quantum nonlinear light-matter interaction.

Our chiral ferroelectric LC platform exhibits distinct advantages over traditional SPDC sources. The self-assembling nature of low molecular-mass organic LCs significantly simplifies the fabrication process, reducing costs and avoiding environmentally harmful chemicals. Its reconfigurability allows for dynamic adjustments of structural parameters (e.g., helical pitch, head and tail director orientations in the helix) via alignment and external stimuli (see thermally and electrically modulations of brightness of entangled photon pairs in **SI Section 6 and Fig. S9**), enabling real-time tuning of quantum photonic properties. In addition, the micro/nano-scale dimension (in $z$-axis, thickness of 50 μm) and macroscopic scalable dimensions ($x$ and $y$-axes) of LC film facilitate seamless integration into quantum photonic circuits, paving the way for miniaturized, high-performance quantum systems. Unlike traditional centimeter-scale nonlinear crystals or even advanced two-dimensional materials[52,53], our platform's micro/nano-scale dimensions provide unparalleled freedom in designing three-dimensional structural engineering. This capability allows spatially precise control over the orientation of individual LC dipoles combined with temporally tuning, influencing the second-order nonlinear susceptibility distribution and thus enabling unprecedented functional flexibility.

This work may open up exciting possibilities for advancing quantum photonics[54]. As essential resources in quantum techniques, the experimental realization of nonreciprocal quantum photon-pair sources is just the first step. By integrating photopatterning techniques with reconfigurable chiral helical structures[46], this platform enables complex functional architectures for manipulating multiple degrees of freedom of twin photons, including path, angular momentum, and spatial modes, promising the creation of hyperentangled states as well as their nonreciprocal control in higher-dimensional Hilbert spaces[55]. This also offers fertile ground for exploring exotic quantum optical phenomena and developing innovative solutions in areas like noise-resilient quantum communication, sensing, and information processing. This study marks a significant step forward in the development of nonreciprocal quantum photon-pair sources, showcasing the exceptional capabilities of chiral ferroelectric LC platforms. The demonstrated achievements in nonreciprocity, efficiency, robustness, and functional flexibility position our platform as a transformative tool for quantum photonics research and applications.

## Acknowledgements

The authors gratefully appreciate Dr. Jie Wang from Hunan Normal University for her constructive discussions.

We acknowledge financial support from National Key Research and Development Program of China (No. 2022YFA1405000 (Y.-Q.L)); National Natural Science Foundation of China (Nos. T2488302 (Y.-Q.L), 62375119 (L.-L.M), T2325022 (X.-F.R.), U23A2074 (X.-F.R.), GG2470000198 (Y.-K.W.), and Grant 12421005 (H.J.)); National Key Research and Development Program of China (Nos. 2021YFA1202000 (L.-L.M), 2022YFA1204704 (X.-F.R.), and 2024YFE0102400 (H.J.)); Natural Science Foundation of Jiangsu Province (Nos. BK20243067 (Y.-Q.L) and BK20232040 (L.-L.M)); Fundamental Research Funds for the Central Universities (Nos. 2024300360 (L.-L.M), WK2030000108 (X.-F.R.), and WK2030000083 (Y.-K.W.)); the Innovation Program for Quantum Science and Technology (Nos. 2021ZD0303200 (X.-F.R.) and 2021ZD0301500 (X.-F.R.)); the Hunan Major Sci-Tech Program (No. 2023ZJ1010 (H.J.)); the CAS Project for Young Scientists in Basic Research (No. YSBR-049 (X.-F.R.)). This work was partially carried out at the USTC Center for Micro and Nanoscale Research and Fabrication.

## Contributions

L.-L.M., X.-F.R. and Y.-Q.L. conceived the basic idea and supervised its development. J.-T.P., Y.-K.W., L.-L.M., N.W., X.-Y.T., and B.-H.Z. implemented the experiment. J.-T.P. and Y.-K.W. provided the simulations and performed the theoretical analysis. J.-T.P., Y.-K.W., L.-L.M., X.-F.R., H.J and Y.-Q.L. wrote the paper with input from all the authors. All the authors discussed the results and contributed to the paper.